\documentclass[aps,prl,twocolumn,superscriptaddress]{revtex4-2}
\usepackage{bbm}
\usepackage{mathrsfs}
\usepackage{amsmath}
\usepackage{amsfonts}
\usepackage[colorlinks=true,linkcolor=blue,urlcolor=blue,citecolor=blue,anchorcolor=blue]{hyperref}
\usepackage{graphicx,epstopdf}
\usepackage{subfigure}
\usepackage{epsfig}
\usepackage{dcolumn}
\usepackage{bm}
\usepackage{color}
\usepackage{natbib}
\usepackage{amssymb}
\usepackage{xcolor}
\usepackage{braket}
\usepackage{ulem}
\usepackage{float}
\usepackage{lipsum}
\usepackage{array}
\usepackage{booktabs}

\begin{document}
\title{Projection measurement of the comb basis through free-electron-photon interactions} 

\author{Zihang Zou}
\address{State Key Laboratory of Artificial Microstructure and Mesoscopic Physics, School of Physics, Frontiers Science Center for Nano-optoelectronics, $\&$ Collaborative Innovation Center of Quantum Matter, Peking University, Beijing 100871, China}
\author{Feng-Xiao Sun}
\email{sunfengxiao@bupt.edu.cn}
\address{State Key Laboratory of Information Photonics and Optical Communications, Beijing Key Laboratory of Information Metamaterials, $\&$ School of Physical Science and Technology, Beijing University of Posts and Telecommunications, Beijing 100876, China}
\address{State Key Laboratory of Artificial Microstructure and Mesoscopic Physics, School of Physics, Frontiers Science Center for Nano-optoelectronics, $\&$ Collaborative Innovation Center of Quantum Matter, Peking University, Beijing 100871, China}
\author{Yunquan Liu}
\address{State Key Laboratory of Artificial Microstructure and Mesoscopic Physics, School of Physics, Frontiers Science Center for Nano-optoelectronics, $\&$ Collaborative Innovation Center of Quantum Matter, Peking University, Beijing 100871, China}
\address{Collaborative Innovation Center of Extreme Optics, Shanxi University, Taiyuan, Shanxi 030006, China}
\author{Qiongyi He}
\email{qiongyihe@pku.edu.cn}
\address{State Key Laboratory of Artificial Microstructure and Mesoscopic Physics, School of Physics, Frontiers Science Center for Nano-optoelectronics, $\&$ Collaborative Innovation Center of Quantum Matter, Peking University, Beijing 100871, China}
\address{Collaborative Innovation Center of Extreme Optics, Shanxi University, Taiyuan, Shanxi 030006, China}
\address{Hefei National Laboratory, Hefei 230088, China}


\begin{abstract}
Free electrons, driven by rapid advances in photon-induced near-field electron microscopy, have emerged as a promising platform for quantum information processing, including quantum computing and quantum sensing. However, conventional measurements that rely on the electron energy loss spectrum (EELS) are inherently destructive to electron qubits, thereby constraining their applicability. In this Letter, we propose a scheme that performs projection measurement on the electron comb basis, where high measurement precision can be achieved with bright squeezed vacuum states and strong PINEM couplings. Notably, this approach is not only nondestructive to electron qubits but also maximally incompatible with energy measurements, enabling alternative quantum information applications, such as quantum error mitigation and Einstein-Podolsky-Rosen steering detection. Our findings open an avenue towards a systematic understanding of quantum free electrons and towards the development of nondestructive free electron quantum information tasks.
\end{abstract} 

\maketitle


In recent decades, rapid experimental advances in ultrafast electron microscopy~\cite{kfir2020controlling,wang2020coherent,garcia2010optical,baum2007attosecond} have facilitated the photon-induced near-field electron microscopy (PINEM) effects~\cite{barwick2009photon,garcia2010multiphoton,park2010photon,Pan2019Anomalous,de2025roadmap}, which reveal the preparation of quantum states~\cite{fang2024structured,Fang2026Electron,fock-prepare,sun2023generating,Huang2023Electron,DiGiulio2022Optical,gkp-prepare} and nonclassical correlations~\cite{schattschneider2018entanglement,Kfir2019Entanglements,Mechel2021Quantum,baranes2022free,Feist2022Cavity} arising from the interaction between free electrons and ultrafast light~\cite{dahan2020resonant,Roques2023Free,Karnieli2023Jaynes,Shi2025Quantum}, and give rise to an emerging field of free-electron quantum optics~\cite{ruimy2025free}. Within this framework, the free electron is commonly regarded as an alternative quantum resource capable of carrying quantum information, with wide applications in ultrafast quantum information processing such as quantum sensing~\cite{FERI,g_ref,ele-photon-ent,turner2021interaction,Chen2023Quantum,Karnieli2023Quantum,bucher2024coherently,shi2024transverse,gorlach2024photonic} and quantum computing~\cite{reinhardt2021free,ele-GKP,free-electron-computing,tsarev2021free}. 

Nevertheless, conventional free-electron measurement schemes remain largely limited to the electron energy loss spectrum (EELS)~\cite{egerton2009electron}, which projects the free electron $\ket \psi_e=\sum_E  c_E \ket E$ to its energy eigenstates $\ket E$ and is therefore destructive to the electron coherence and the free-electron qubit states. Consequently, quantum tasks including quantum error mitigation~\cite{cai2023quantum,endo2018practical} that require nondestructive measurements~\cite{grangier1998quantum}, remain infeasible with current techniques. Furthermore, the lack of measurement schemes to directly access the phase information of free electron states, which is maximally incompatible with the electron energy spectrum, restricts the realization of their full quantum characteristics~\cite{wootters1989optimal,MAURODARIANO2003205,skrzypczyk2019all}. For instance, certain quantum correlation verification cannot be achieved because it requires incompatible measurements~\cite{uola2020quantum,reid1989demonstration,EPR-Reid-2009,duan2000inseparability}. Thus, in order to extend the application of free electrons in quantum tasks, it is crucial to develop an alternative measurement that is both incompatible with EELS and nondestructive to the electron coherence.

In this Letter, we propose a scheme to perform free electron comb-basis projection measurement (FECBPM) to fill this gap. In our scheme, a free electron passes through a PINEM interaction and becomes entangled with an optical mode. Subsequently, through homodyne detection on the optical state, the electron is projected to the corresponding comb state $\ket {\text{comb}(\phi)}\propto \sum_k e^{-ik\phi}\ket{E-\hbar k\Omega}$ with the phase $\phi\in[-\pi,\pi)$. In this way, FECBPM can be performed non-destructively on the electron state, as the homodyne detection is applied only to the auxiliary optical states. Furthermore, we show that the nondestructive feature of FECBPM enables quantum error mitigation for free-electron qubits. Meanwhile, the fact that the comb basis is the Fourier transform of the electron energy basis implies that FECBPM possesses quantum incompatibility with EELS~\cite{quant-MUB}, which can be recognized as an essential resource for various quantum information tasks~\cite{incompat-resource,multi-para-incompat}. Here, we demonstrate its application in the detection of Einstein-Podolsky-Rosen (EPR) steering~\cite{Wiseman-2007,Reid2009Colloquium,EPR-Xiang,He2013Genuine,XIANG2021Advances,tian2024certification} between the free electron and the optical mode by combining FECBPM with EELS.


Now, in order to perform FECBPM, we refer to the PINEM interaction between the free electron and an optical mode, which is described by the unitary operator,
\begin{equation}
    U(g)=\exp(g a_{ph}^\dagger b_\Omega-H.c.),
\end{equation}
where $g$ is the coupling strength, $a_{ph}$ is the photon annihilation operator with frequency $\Omega$, $b_\Omega$ is the electron energy ladder operator satisfying $b_\Omega\ket {E}=\ket{E-\hbar \Omega}$ and $b_\Omega^\dagger b_\Omega=b_\Omega b_\Omega^\dagger=1$, whose eigenstate is the free electron comb state with $b_\Omega\ket {\text{comb}(\phi)}=e^{i\phi}\ket{\text{comb}(\phi)}$ and $\langle \text{comb}(\phi')\ket {\text{comb}(\phi)}=\delta_{\phi\phi'} $. Although free-electron comb states are of significant theoretical interest due to their status as eigenstates of the ladder operator $b_\Omega$, their direct measurement remains a substantial challenge in experiments.

The effects of the PINEM interaction between a free electron comb state $\ket {\text{comb}(\phi)}$ and an arbitrary photon state $\ket{\psi_{ph}}$ can be directly obtained,
\begin{equation}
    U(g)[\ket {\text{comb}(\phi)}\otimes \ket {\psi_{ph}}]=\ket{\text{comb}(\phi)}\otimes D(g e^{i\phi})\ket{\psi_{ph}},
    \label{eq:displacement}
\end{equation}
which is equivalent to a displacement operator $D(g e^{i\phi})=\exp(g e^{i\phi}a_{ph}^\dagger-g^\ast e^{-i\phi}a_{ph})$ performed on the photon state. That is to say, electrons in different comb states displace the photon in different directions through PINEM interactions. This motivates us to measure the auxiliary optical mode so as to read the phase of the free electron comb state non-destructively.  

\begin{figure}[t!]
    \centering
    \includegraphics[width=\linewidth]{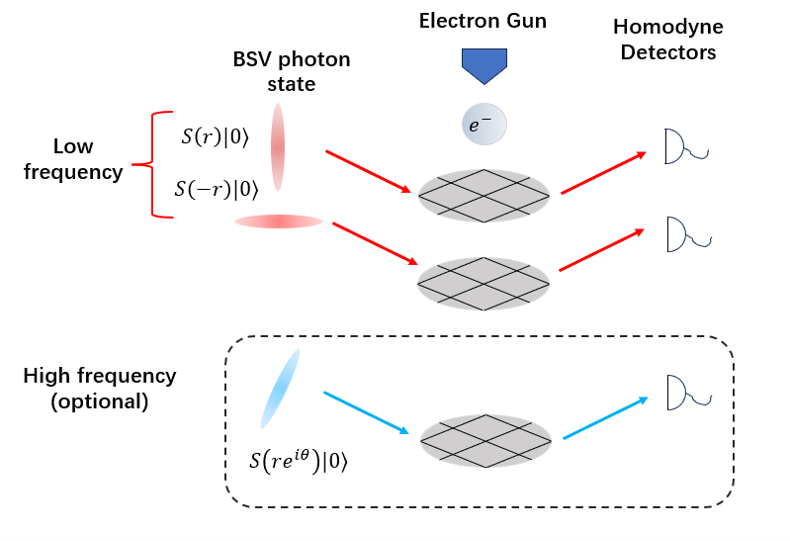}
    \caption{The experimental scheme for FECBPM. The low frequency part is based on the PINEM platform, where free electron interacts with two optical modes of orthogonal squeezing angles. Then we perform homodyne detections on the output optical modes, and the outcomes project the free electron to the associated comb state non-destructively. Further, the high frequency part can be applied to enhance the measurement accuracy.}
    \label{apparatus}
\end{figure}

To perform the projective measurement, the photon state after the PINEM interaction is required to be distinguishable so that the state $D(g e^{i\phi})\ket{\psi_{ph}}$ should be approximately orthogonal for different $\phi$s. For an ideal case, the free electron with initial state $\sum c_\phi\ket{\text{comb}(\phi)}$ goes through PINEM interaction with two photon modes $1$ and $2$, prepared in eigenstates of $X_1=a_1^\dagger +a_1$ and $P_2=i(a_2^\dagger-a_2)$ with eigenvalue $0$, respectively. Thus, the system state is 
\begin{equation}
    \ket{\psi_0}=\sum c_\phi\ket{\text{comb}(\phi)}\otimes \ket{X_1=0}\otimes \ket{P_2=0}.
\end{equation}
Setting the coupling strengths of the two PINEMs to be $g_1$ and $g_2$ (both real numbers for simplicity), the final state after the interaction becomes 
\begin{align}
    \ket{\psi}=\sum c_\phi\ket { \text{comb}(\phi)}\otimes \ket{X_1=g_1\cos\phi}&\otimes \ket{P_2=g_2\sin\phi}.
\end{align}
Therefore, when we implement homodyne measurement to measure $X_1$ and $P_2$ of the photon modes, with the outcomes $x_1$ and $p_2$ obtained respectively, the electron state is post selected to $\phi=\arctan(x_1g_2/g_1p_2)$ if $x_1$ and $g_1$ have the same sign, or to $\phi=\arctan(x_1g_2/g_1p_2)+\pi$ if $x_1$ and $g_1$ have the opposite sign. 

Considering the experimental feasibility, the eigenstates of $X$ and $P$ can be replaced with bright squeezed vacuum states. As shown in Fig.\ref{apparatus}, two squeezed states with orthogonal squeezed angles $S(r)\ket 0$ and $S(-r)\ket 0$ are used as auxiliary photon modes, where $S(r)=\exp[r(a^2-a^{\dagger2})/2]$ with the squeezing parameter $r$. Recently, the squeezing parameter approaching $r\sim15$ can be achieved with state-of-the-art techniques~\cite{Rasputnyi2024High,Heimerl2025Quantum,Lyu2026Generation}. In the limit of large squeezing $r$, the squeezed states become increasingly indistinguishable from the eigenstates of the corresponding quadratures, offering promise for realizing near-perfect FECBPM. To further improve the measurement precision under practical conditions with finite squeezing, we can adopt the high-frequency part shown in Fig.\ref{apparatus} (see details in the Supplemental Material (SM)~\cite{supplementary}). This is experimentally feasible as the PINEM effects have been reported with the optical frequency ranging from $290$THz to $580$THz~\cite{290Hz}.

\begin{figure}[t]
    \centering
  \hspace*{-1.0cm}
    \includegraphics[width=1.0\linewidth]{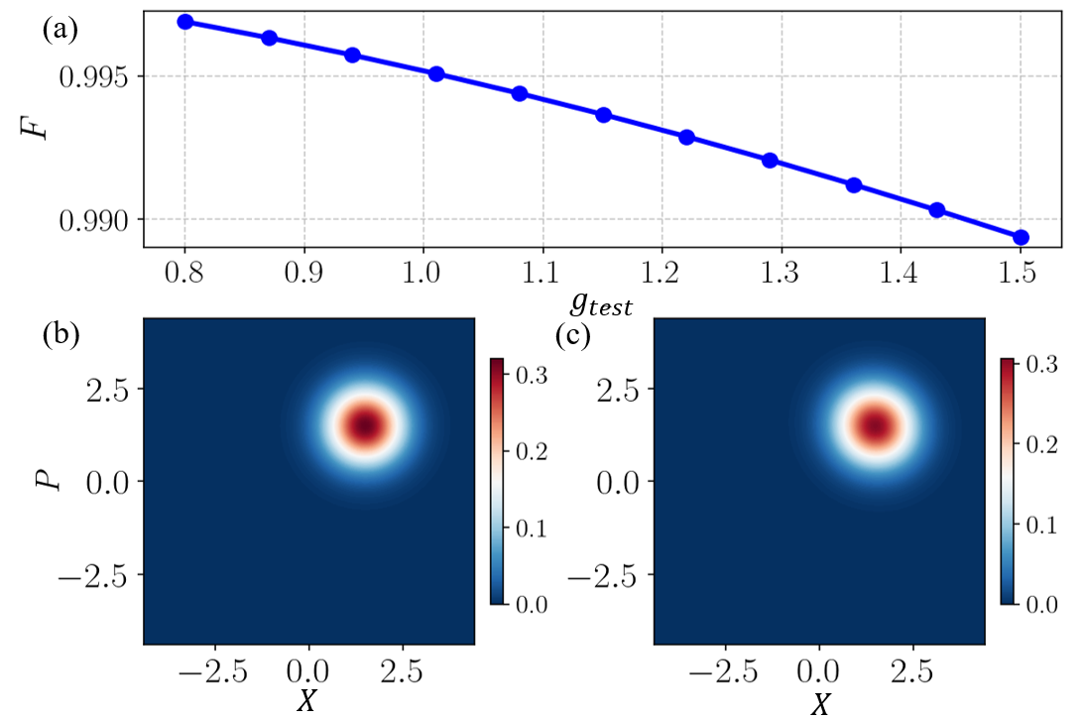}
     \caption{(a) The fidelity between the auxiliary optical states generated from the electron obtained in our FECBPM scheme $\ket{\psi_{ph}^{out}}$ and the ideal comb state $\ket{\psi_{ph}^{ideal}} $ for different coupling strengths $g_{test}$. And the Wigner functions of (b) $\ket{\psi_{ph}^{out}}$ and (c) $\ket{\psi_{ph}^{ideal}}$ are present with $g_{test}=1.0$, respectively. Other parameters are $g=\sqrt{2}$, $r=1.5$ and $\phi=\pi/4$.}
      \label{state-compare}
\end{figure}

In order to evaluate our FECBPM scheme, we need to compare the electron output state $\ket{\psi_{e}^{out}}$ and the ideal state $\ket{\psi_{e}^{ideal}}$. However, due to the continuous-variable $\phi$ and finite squeezing in our FECBPM scheme, the fidelity of the electron comb states is no longer a reliable figure of merit to quantify their similarity. In particular, the fidelity between $\ket{\text{comb}(\phi)}$ and $\ket{\text{comb}(\phi+\delta\phi)}$ remains $0$ even for $\delta\phi\ll\pi$, despite the fact that these two states are very close to each other. Thus, we will refer to the fidelity between the optical modes, which are generated via the PINEM interaction between the vacuum state and the obtained/ideal electron state. Note that the electron-photon state after the PINEM interaction is in the form of $\ket{\psi_{e-ph}^{i}} =  \exp(g_{test} a_{ph}^\dagger b_\Omega-H.c.)\ket{\psi_{e}^{i}}\ket{0_{ph}}$ with $i=out,\,ideal$, the reduced optical state is then ${\rho_{ph}^{i}}=\text{Tr}_e(\ket{\psi_{e-ph}^{i}}\bra{\psi_{e-ph}^{i}})$. By analyzing the fidelity of the optical states generated by $\ket{\text{comb}(\phi)}$ and $\ket{\text{comb}(\phi+\delta\phi)}$, the minimum distinguishable $\delta \phi $ is proportional to $1/ |g| e^{r}$ if $g_1=g_2=g$ (See details in SM~\cite{supplementary}). Considering a case where the electron is projected from the energy eigenstate $\ket 0_e$ to the comb state $\ket{\text{comb}(\pi/4)}$ by applying FECBPM with the squeezing parameter $r=1.5$ and the PINEM couplings $g=\sqrt 2$, the fidelity is then illustrated changing with $g_{test}$ in Fig.~\ref{state-compare}(a). Taking $g_{test} =1.0$ as an example, the fidelity of the obtained state and the ideal state is above $99.5\%$, whose Wigner functions can be found in Fig.~\ref{state-compare}(b) and (c), respectively.

We further discuss the potential applications of our proposed FECBPM in quantum information tasks. The first is about fault-tolerant quantum computation~\cite{FTQC-nature,FTQC,QEC}. Within the framework of PINEM, free electrons have been proposed as an alternative platform to construct qubits~\cite{reinhardt2021free}, demonstrating their potential for quantum computing and quantum manipulation~\cite{ele-GKP,free-electron-computing}. However, the quantum errors of free-electron qubits in quantum computation tasks have not been discussed yet. Here, by exploiting the non-destructive property of FECBPM, we propose a quantum error mitigation scheme that suppresses errors arising from free-space propagation (FSP).

The free electron can be encoded as a qubit, with logical bits $\ket 0_L=\sum_{k=-\infty}^{+\infty} \ket{E-2k\hbar \Omega}=\frac 1{\sqrt 2}(\ket {\text{comb}(0)}+\ket {\text{comb}(\pi)})$ and $\ket{1}_L = \sum_{k=-\infty}^{+\infty} \ket{E-(2k-1)\hbar\Omega} = \frac{1}{\sqrt{2}}(\ket{\text{comb}(0)} - \ket{\text{comb}(\pi)})$,  where $\Omega$ is the frequency of light in the free electron qubit circuit.
These free electron qubits can be manipulated by the PINEM interaction and the FSP. In particular, the FSP introduces a phase factor to each energy component,
\begin{equation}
    F(\phi)=\sum_k \exp(-ik^2\phi)\ket{E-k\hbar \Omega}\bra{E-k\hbar \Omega},
\end{equation}
where $\phi$ is proportional to the propagation distance. Specially, the Pauli-Z gate is achieved with $F(\phi=\frac\pi2)$. However, errors in the propagation distance will drive the electron state out of the code space spanned by $\ket 0_L$ and $\ket 1_L$, indicating the necessity of quantum error mitigation.

We notice that $\ket 0_L$ and $\ket 1_L$ are both eigenstates of $b_{2\Omega}=b_\Omega^2$ with eigenvalue $1$, which leads $b_{2\Omega}$ to be a stabilizer. This means that, by the PINEM interaction with the coupling strength $g$, both $\ket 0_L$ and $\ket 1_L$ effectively perform a displacement $D(g)$ to the optical mode of frequency $2\Omega$. Hence, the stabilizer measurement scheme can be performed by applying the FECBPM with optical frequency $2\Omega$. It projects the electron state to the subspace $\{\ket{\text{comb}(\phi)} ,\ket{\text{comb}(\phi+\pi)} \}$, where the eigenvalue of the stabilizer $b_{2\Omega}$ is $e^{2i\phi}$ with $\phi\in[0,\pi)$. Thus, the free electron qubit is verified in the code space when $\phi=0$ is measured. Nevertheless, if $\phi\ne 0$ is detected, it necessarily signals the occurrence of errors, which can be corrected in subsequent gates of the free‑electron quantum computing circuits by adjusting $g\to g e^{-i\phi}$ (See SM~\cite{supplementary}).

\begin{figure}
    \centering
    \includegraphics[width=1.0\linewidth]{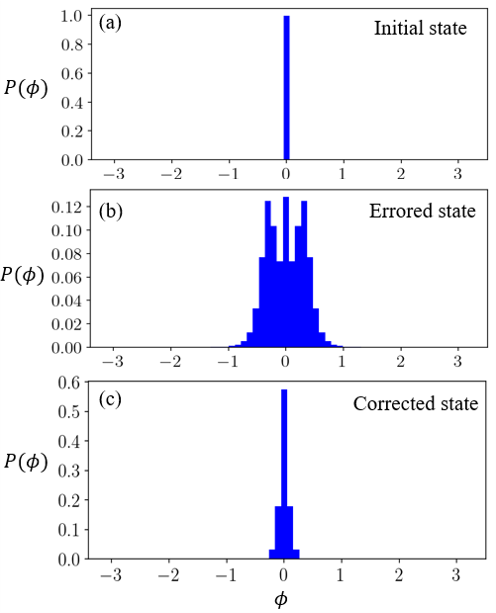}
    \caption{The probability distribution on comb basis of (a) the initial state, (b) the state after the FSP with errors $\delta=0.01$, and (c) the state after error mitigation with $g=\sqrt{2}$ and $r=1.5$.}
    \label{fig:pro_dis}
\end{figure}

Here, we assume that the initial state is $\ket{\psi_e^{(0)}}=\frac 1{\sqrt N}\sum_{k=-\infty}^{+\infty} e^{-\frac{k^2}{2\sigma^2}}  \ket {E-k\hbar \Omega}$ with variance  $\sigma=50$, which is a practical approximation for $\ket 0_L$, whose probability distribution on comb basis is shown in Fig.~\ref{fig:pro_dis}(a). Then errors are included through the FSP of the electron, resulting in 
\begin{equation}
    \ket{\psi_e^{(error)}}=\frac 1{\sqrt N}\sum_{k=-\infty}^{+\infty} e^{-\frac{k^2}{2\sigma^2}-ik^2\delta}  \ket {E-k\hbar \Omega},
\end{equation}
where $\delta$ is proportional to the FSP distance errors. By truncating the electron Hilbert space to $N_d=64$ (with $k$ ranging from$-32$ to $31$) and setting $\delta=0.01$, the distribution is therefore significantly changed [Fig.~\ref{fig:pro_dis}(b)]. To suppress the effects of the FSP errors, we take $g=\sqrt 2$  in our quantum error mitigation scheme, such that the corrected state by adjusting the gates approximately projects into the code space $\{\ket{\text{comb}(0)} ,\ket{\text{comb}(\pi)} \}$, with its probability distribution plotted in Fig.~\ref{fig:pro_dis}(c). It is seen that our scheme approximately reconstructs the initial distribution, demonstrating its capability to mitigate the FSP errors of the free electron qubit.

Subsequently, we will introduce another application of our FECBPM in quantum correlation detection. In previous work, the quantum correlation between the electron and the photon was described simply by energy correlation, as only energy measurement can be applied to the electrons~\cite{ele-photon-ent}. Here, we focus on EPR steering, which is a proper subset of quantum entanglement where local measurement on one party can remotely affect the state of another party significantly, violating the local hidden state model \cite{Wiseman-2007,Reid2009Colloquium,EPR-Xiang}. 
To witness EPR steering, incompatible measurements are required based on the Reid criterion~\cite{reid1989demonstration,EPR-Reid-2009}, which indicates that the system $B$ can steer the system $A$ if the following inequality is violated,
\begin{equation}\label{eq:EPR}
     \Delta^2_{inf}X_A\Delta_{inf}^2 P_A\ge\frac 14|\langle [X_A,P_A]\rangle|^2.
 \end{equation}
Here, the operators $X_A$, $P_A$ ($X_B$, $P_B$) are arbitrary operators in the system $A$ ($B$), where the inferred variance $\Delta^2_{inf}X_A=\int dx_B \Delta^2(X_A|X_B=x_B) P(X_B=x_B)$, $\Delta^2(X_A|X_B=x_B)$ is the variance of $X_A$ when $X_B=x_B$, and $P(X_B=x_B)$ is the probability of $X_B=x_B$. Similarly, we define $\Delta^2_{inf}P_A=\int dp_B \Delta^2(P_A|P_B=p_B) P(P_B=p_B)$. In order to violate Eq.~(\ref{eq:EPR}), the right-hand term is expected to be as large as possible, indicating $X_A$ and $P_A$ to be maximally incompatible. In the following, we will take two examples to show that EPR steering between the electron and the photon can be verified by combining FECBPM and EELS.

\begin{figure}
    \centering
    \hspace*{-0.5cm}
    \includegraphics[width=1.10\linewidth]{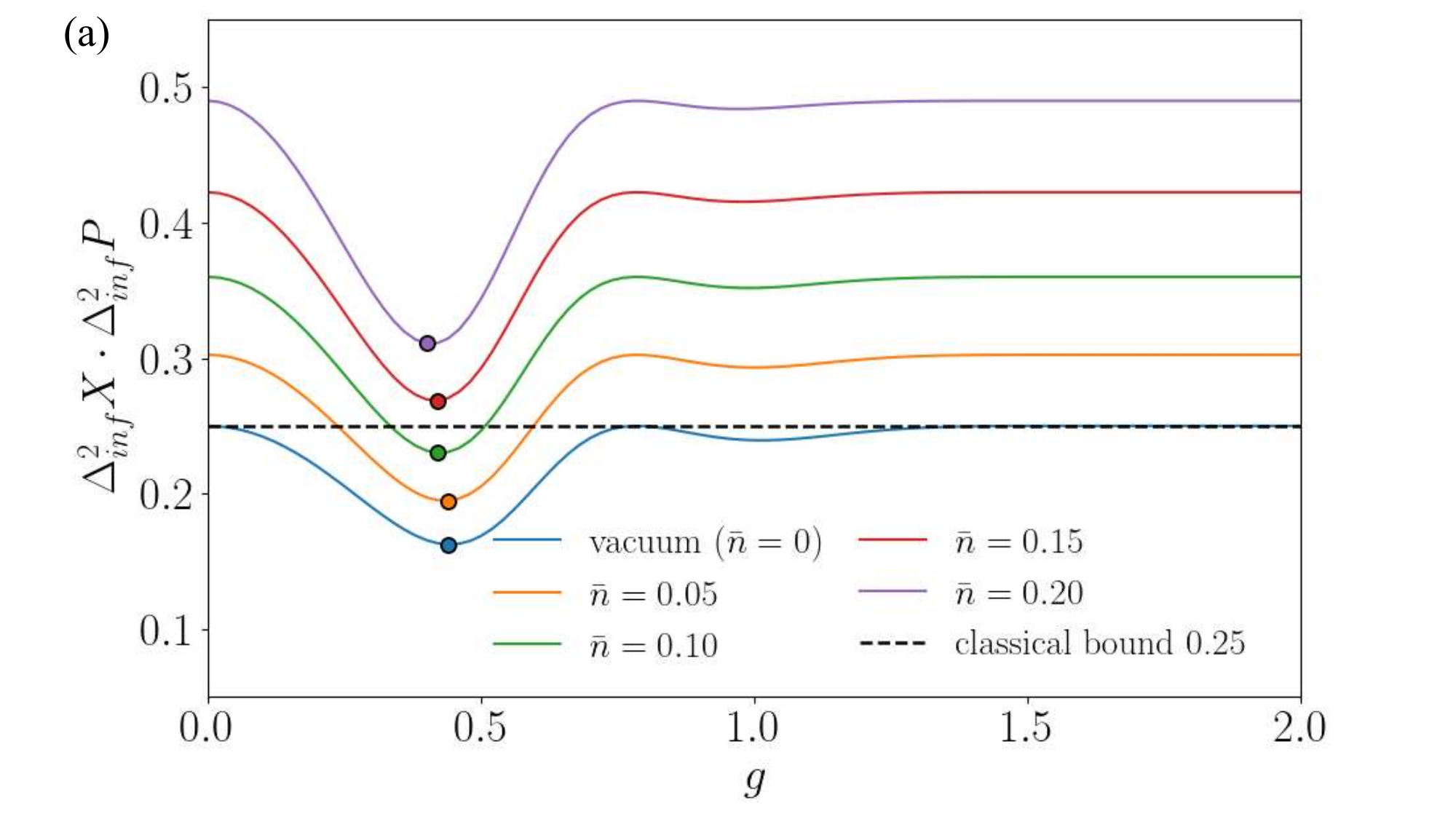}
    \hspace*{-0.49cm}
    \includegraphics[width=0.95\linewidth]{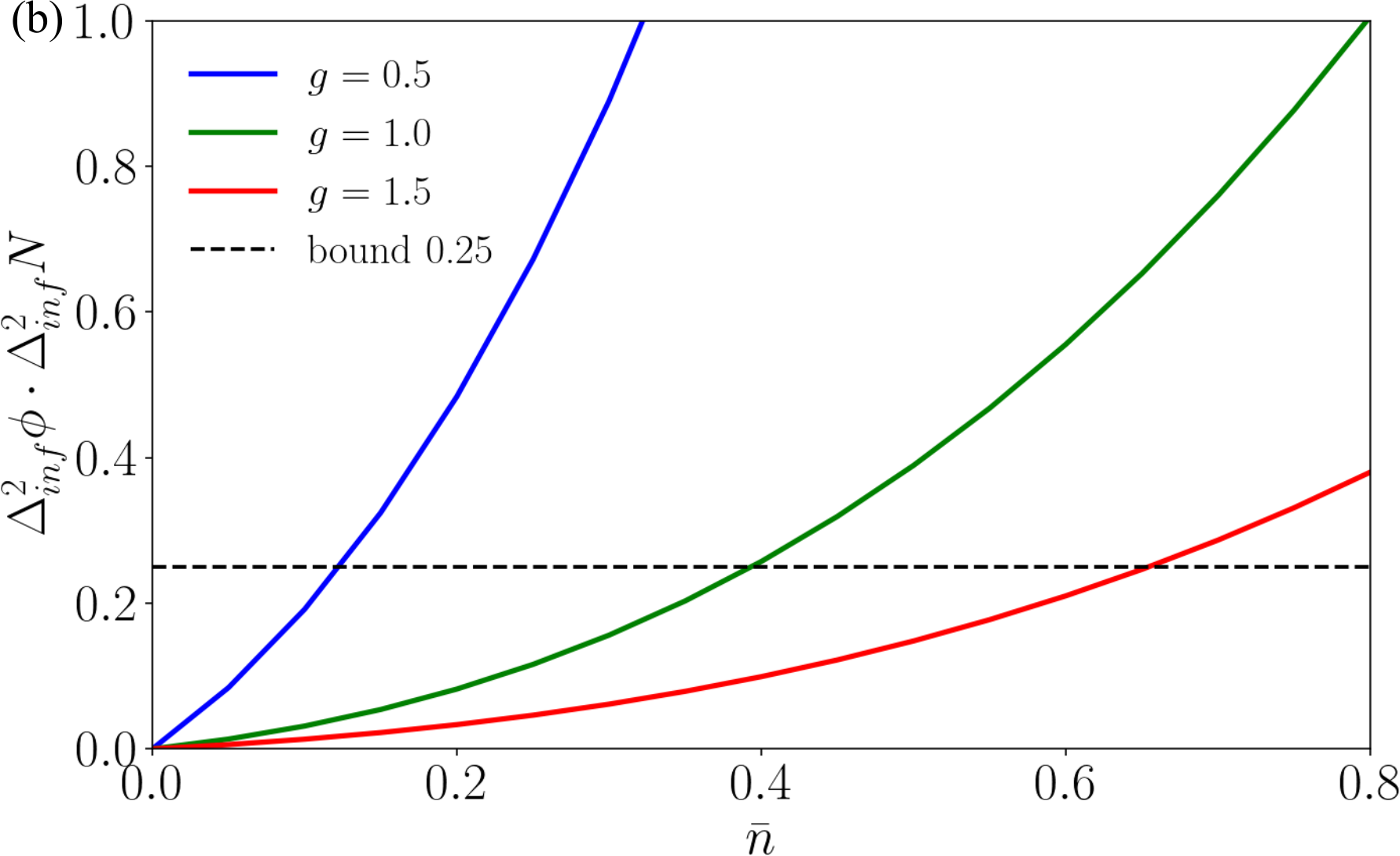}
    \caption{EPR steering detection for electron-photon states. (a) The values of $\Delta ^2_{inf} X\Delta^2_{inf}P$ for coupled electron qubits and optical displaced thermal states [Eq.~(\ref{ent-state-xp})] varies with the coupling strength $g$ for different average photon numbers $\bar n$. (b) $\Delta ^2_{inf} N\Delta^2_{inf}\phi$ varies with the average photon number $\bar n$ of the optical thermal states for different PINEM coupling strengths $g$. The black dashed line denotes the EPR steering bound, below which EPR steering can be detected.    }
    \label{steering}
\end{figure}

As the first example, we consider the coupling between the free electron qubit and the optical mode, where the initial electron state is $\ket 0_L$ and the initial optical state is a coherent state $\ket{i\alpha}$. After the PINEM interaction, the electron-photon state becomes $\frac1{\sqrt 2}[D(g)\ket{i\alpha}\ket {\text{comb}(0)}+D(-g)\ket{i\alpha}\ket {\text{comb}(\pi)}]$. For simplicity, $g$ and $\alpha$ are set as real numbers. Here, we choose quadrature operators for the optical mode $X_A=\frac 1{\sqrt 2}(a+a^\dagger)$, $P_A=\frac i{\sqrt 2}(a^\dagger-a)$, and Pauli operators for the free electron qubit $X_B=\sigma_z$ (corresponding to the energy measurement with EELS) and $P_B=\sigma_x$ (corresponding to FECBPM). Hence, $\Delta^2_{inf}X_A\Delta_{inf}^2 P_A$ with $\alpha=1.0$ is shown in Fig.~\ref{steering}(a) with the solid blue line, indicating that the EPR steering can be detected for any $g$ and $\alpha$ (See the proof in SM~\cite{supplementary}). We further generalize the photon initial state to a displaced thermal state $\rho_{ph}=D(i\alpha)\sum_n\frac{\bar n ^n\ket n \bra n}{(1+\bar n)^{n+1}}  D(-i\alpha)$ with the average thermal photon number $\bar n=(e^{\hbar\omega/k_BT}-1)^{-1}$, where $T$ is the temperature and $k_B$ is the Boltzmann constant. The electron-photon state after the PINEM interaction is 
\begin{equation}
\label{ent-state-xp}
    \rho_{ele-ph}=U(g)[\rho_{ph}\otimes \ket0_L\bra{0}_L]U^\dagger(g).
\end{equation}
We present the values of $\Delta^2_{inf}X_A\Delta_{inf}^2 P_A$ for different $\bar n$s in Fig.~\ref{steering}(a), revealing that the EPR steering can be detected for at least $\bar n<0.1$ with proper $g$.

In the second example, we consider the initial state $\ket 0_{ph}\ket 0_e$ to be an optical vacuum state with an energy-confined electron, so that the state after a PINEM interaction is $\sum c_k \ket k_{ph}\ket{-k}_e$. Then we choose the photon number and phase operators for the optical mode, $X_A=N_{ph}$, $P_A=\phi_{ph}$, which satisfy $[N_{ph},\phi_{ph}]=i$, and the energy and phase operators for the electron, $X_B=E_e$, $P_B=\phi_e$ (corresponding to EELS and FECBPM respectively). For this ideal case, the conditional state after electron energy measurement with EELS is a Fock state, with $\Delta^2N_{ph}=0$, and the conditional state after FECBPM is a coherent state, with $\Delta^2\phi_{ph}<+\infty$. Therefore, we have $\Delta^2_{inf} N_{ph}\Delta^2_{inf} \phi_{ph}=0<1/4$, indicating the electron can always steer the light.
For more practical cases, the initial optical state is generalized to the thermal state $\rho_{ph}^{th}=\sum_{n=0}^{\infty} \frac{\bar{n}^{n}}{(\bar{n} + 1)^{n+1}} \, |n\rangle\langle n|$. Figure~\ref{steering}(b) displays the dependence of $\Delta ^2_{inf} N\Delta^2_{inf}\phi$ on the average photon numbers $\bar{n}$ of the initial thermal optical states, for various PINEM coupling strengths $g$. The EPR steering is observed for sufficiently small $\bar{n}$, and we reveal that the robustness of EPR steering against thermal noise $\bar n$ increases with $g$.


To conclude, we propose a scheme to realize comb-basis projection measurements on free electrons. In our scheme, the free electron is entangled with auxiliary squeezed optical modes via PINEM interaction and is subsequently projected to the corresponding comb basis through homodyne detection on the optical modes. By increasing the PINEM interactions and the squeezing parameter, or involving high-frequency optical modes, arbitrary measurement precision can be achieved. The performance of our scheme is evaluated by comparing the fidelities of the coupled optical modes, showing that the obtained electron quantum state closely approximates the ideal state. In addition, we demonstrate the capability of our proposed FECBPM in specific quantum information tasks. For instance, due to its nondestructive nature, FECBPM enables stabilizer measurements for free-electron qubits and successfully mitigates the FSP error. Moreover, the incompatibility between FECBPM and EELS allows us to witness the EPR steering between the free electron and the optical mode. Our findings pave the way towards extending applications in ultrafast quantum information processing with nondestructive manipulation and measurement on free electrons.

\begin{acknowledgments}
This work is supported by the National Natural Science Foundation of China (Grant Nos. 12474256, 12125402, 12334013, 12534016 and 92250306), the Quantum Science and Technology-National Science and Technology Major Project (Grant Nos. 2025ZD0301000 and 2024ZD0302401), the State Key R\&D Program (Grant No. 2022YFA1604301), and Beijing Natural Science Foundation (Grant No. Z240007). F.-X. S. acknowledges support from the Fundamental Research Funds for the Central Universities (Grant No. 539926010).
\end{acknowledgments}

\bibliography{ref}

\end{document}


\title{Supplemental Material for ``Projection measurement of the comb basis through free-electron-photon interactions''} 

\author{Zihang Zou}
\address{State Key Laboratory of Artificial Microstructure and Mesoscopic Physics, School of Physics, Frontiers Science Center for Nano-optoelectronics, $\&$ Collaborative Innovation Center of Quantum Matter, Peking University, Beijing 100871, China}
\author{Feng-Xiao Sun}
\email{sunfengxiao@bupt.edu.cn}
\address{State Key Laboratory of Information Photonics and Optical Communications, Beijing Key Laboratory of Information Metamaterials, $\&$ School of Physical Science and Technology, Beijing University of Posts and Telecommunications, Beijing 100876, China}
\address{State Key Laboratory of Artificial Microstructure and Mesoscopic Physics, School of Physics, Frontiers Science Center for Nano-optoelectronics, $\&$ Collaborative Innovation Center of Quantum Matter, Peking University, Beijing 100871, China}
\author{Yunquan Liu}
\address{State Key Laboratory of Artificial Microstructure and Mesoscopic Physics, School of Physics, Frontiers Science Center for Nano-optoelectronics, $\&$ Collaborative Innovation Center of Quantum Matter, Peking University, Beijing 100871, China}
\address{Collaborative Innovation Center of Extreme Optics, Shanxi University, Taiyuan, Shanxi 030006, China}
\address{Peking University Yangtze Delta Institute of Optoelectronics, Nantong 226010, Jiangsu, China}
\author{Qiongyi He}
\email{qiongyihe@pku.edu.cn}
\address{State Key Laboratory of Artificial Microstructure and Mesoscopic Physics, School of Physics, Frontiers Science Center for Nano-optoelectronics, $\&$ Collaborative Innovation Center of Quantum Matter, Peking University, Beijing 100871, China}
\address{Collaborative Innovation Center of Extreme Optics, Shanxi University, Taiyuan, Shanxi 030006, China}
\address{Hefei National Laboratory, Hefei 230088, China}


\maketitle
\tableofcontents

\section{Interaction between electron comb state and optical mode}
In photon-induced near-field electron microscopy (PINEM) systems, the Hamilton for a free electron interacting with photon modes takes the form of~\cite{fock-prepare}
\begin{equation}
    H_{e-ph}=\sum_n \hbar \omega_n a_n^\dagger a_n+\hbar v\cdot p +\sum_n ev[A_z(z) a_n+A_z^\dagger(z) a^\dagger ].
\end{equation}
Here, $a_n$ ($a_n^\dagger$) is the annihilation (creation) operator of the photon mode with frequency $\omega_n$, $A_z(z)$ is the vector potential for the photon mode, $p$ is the electron momentum, and the electron velocity $v$ is considered as a constant. In the interaction picture, the Hamilton becomes 
\begin{equation}
    H_I=\sum_n  W_n(z+vt) a_n e^{-i\omega_n t}+H.c. ,
\end{equation}
where $W_n(z)=ev A_z(z)$. Thus, the scattering operator for the electron-photon interaction is defined as 
\begin{equation}
    U=\mathcal{T}\exp\left(-\frac i\hbar\int_{-\infty}^{+\infty} H_I dt\right)= e^{\chi}\exp\left[\sum_n \beta_n(z) a^\dagger-\beta^*_n(z) a \right],
\end{equation}
where $\chi$ is a classical number and
\begin{equation}
    \beta_n(z)= \left[\frac e{\hbar \omega_n}\int_{-\infty}^{\infty} E_z(z') e^{-i\frac {\omega_n} v z'}dz'\right] e^{-i\frac{ \omega_n} v z}=g_n e^{-i\frac{ \omega_n} v z}.
\end{equation}
Here, $g_n=\frac e{\hbar \omega_n}\int_{-\infty}^{\infty} E_z(z') e^{-i\frac {\omega_n} v z'}dz'$ is the coupling parameter for the photon mode $a_n$. And the term $e^{-i\frac{ \omega_n} v z}$ is the ladder operator of the electron momentum (energy), which can be rewritten as $b_{\omega_n}=\sum_E \ket{E}\bra{E+\hbar \omega_n} $ in the energy basis. Thus, the scattering operator can be rewritten as $U=e^{\chi}\exp(\sum_n g_n a^\dagger b_{\omega_n}-g^*_n a b^\dagger_{\omega_n})$.

Further, we consider the case where the free electron interacts with an optical mode with frequency $m\Omega$, where $m$ is an integer. Thus, if the electron absorbs (emits) such a single high-frequency photon, the energy of the electron will increase (decrease) $m \hbar \Omega$. Therefore, the interaction operator can be expressed as $U(g)=\exp(ga_{ph}^\dagger \sum_E \ket{E}\bra{E+m\hbar \Omega}-H.c. )=\exp(g a_{ph}^\dagger b_\Omega^{m}-H.c.)$, which leads to the relation of $b_{m\Omega}=b_{\Omega}^{m}$.

Here, we assume that the initial electron state is the comb state, $\ket{\text{comb}(\phi)}$. Thus, for arbitrary optical state $\ket{\psi_{ph}}$ with frequency $\Omega$, the photon-electron state after the PINEM interaction becomes 
\begin{equation}
    \ket{\psi_{e-ph}^{(\Omega)}}= \exp(ga^\dagger_{ph}b_\Omega-H.c.)[\ket{\text{comb}(\phi)}\otimes \ket{\psi_{ph}}].
\end{equation}
Noticing that the comb state satisfies $b_\Omega\ket{\text{comb}(\phi)}=e^{i\phi}\ket{\text{comb}(\phi)}$ and $b_\Omega^\dagger \ket{\text{comb}(\phi)}=e^{-i\phi}\ket{\text{comb}(\phi)}$, we have
\begin{equation}
    \ket{\psi_{e-ph}^{(\Omega)}}=\exp(g e^{i\phi} a_{ph}^\dagger-H.c.)[\ket{\text{comb}(\phi)}\otimes \ket{\psi_{ph}}]=\ket{\text{comb}(\phi)}\otimes D(g e^{i\phi})\ket{\psi_{ph}},
\end{equation}
where $D(g e^{i\phi})=\exp(g e^{i\phi}a_{ph}^\dagger-g^\ast e^{-i\phi}a_{ph})$ is the displacement operator. 
And if the free electron comb state interacts with a high-frequency $m\Omega$ optical mode through PINEM, the final state is directly obtained with the form of
\begin{equation}
    \ket{\psi_{e-ph}^{(m\Omega)}}=\ket{\text{comb}(\phi)}\otimes D(g e^{m i\phi})\ket{\psi_{ph}}.
\end{equation}

\section{Condition for reliable projection measurement}
As mentioned in the main text, the state after electron-photon interaction is 
\begin{equation}
    \ket{\psi}=\sum c_\phi\ket{\text{comb}(\phi)}\otimes D(g_1 e^{i\phi})S(r) \ket{0}\otimes D(g_2 e^{i\phi})S(-r) \ket{0}.
\end{equation}
To perform the projection measurement, we require the displaced photon state of different $\phi$s to be distinguishable. As a result, in order to distinguish the states $\ket{\text{comb}(\phi)} $ and $\ket{\text{comb}(\phi+\delta \phi)} $, we require the states $D(g_1 e^{i\phi})S(r) \ket{0}\otimes D(g_2 e^{i\phi})S(-r) \ket{0}$ and $D(g_1 e^{i(\phi+\delta \phi)})S(r) \ket{0}\otimes D(g_2 e^{i(\phi+\delta \phi)})S(-r) \ket{0}$ to be approximately orthogonal, whose fidelity is approximately obtained as
\begin{equation}
    F= \exp\left[-{(\text{Im}(g_1 e^{i\phi})^2+\text{Re}(g_2 e^{i\phi}) ^2)\delta\phi^2e^{2r}}\right].
\end{equation}
With the case considered $g_1=g_2=g$, it can be further simplified to $F=\exp(-|g|^2\delta\phi^2 e^{2r})$. Therefore, the minimal distinguishable $\delta \phi $ is proportional to $1/ |g| e^{r}$. This means that large PINEM coupling strengths and squeezing parameters are required for high precision measurement of the electron comb states.

In realistic experiments, the attainable PINEM coupling strengths and squeezing parameters are generally limited by technical and practical constraints. To fill this gap, we further proposed employing high-frequency light to interact with the free electron. For the initial photon state $\ket{\psi_{ph}^{m\Omega}} $ with frequency $m\Omega$, the final state after the PINEM interaction with coupling strength $g$ is 
\begin{equation}
     \ket{\psi}=\sum c_\phi\ket{\text{comb}(\phi)}\otimes D(ge^{mi\phi}) \ket{\psi_{ph}^{m\Omega}}.
\end{equation}
Therefore, the condition for distinguishing the states $\ket{\text{comb}(\phi)} $ and $\ket{\text{comb}(\phi+\delta \phi)} $ is that the states $ D(ge^{mi\phi}) \ket{\psi_{ph}^{m\Omega}}$ and $D(ge^{mi(\phi+\delta \phi)}) \ket{\psi_{ph}^{m\Omega}}$ are approximately orthogonal. Assuming that high-frequency light is a squeezed vacuum state with squeezing parameter $r$, the fidelity between $ D(ge^{mi\phi}) \ket{\psi_{ph}^{m\Omega}}$ and $D(ge^{mi(\phi+\delta \phi)}) \ket{\psi_{ph}^{m\Omega}}$ is $F=\exp(-m^2|g|^2\delta\phi^2 e^{2r})$, so that the minimal distinguishable $\delta\phi  $ is proportional to $1/m |g| e^{r}$. Therefore, the measurement precision is improved by using high-frequency optical squeezed state, even if the PINEM coupling strengths and the squeezing parameters are limited.

\section{Homodyne detection on the auxiliary optical mode for free electron qubit}
Here we consider a special case where the free electron state is confined to a 2-dimensional Hilbert space spanned by $\ket {\text{comb}(0)}$ and $\ket {\text{comb}(\pi)}$, which can be regarded as a free electron qubit. In this case, only using one single squeezed vacuum state as the auxiliary optical mode is sufficient to perform the FECBPM, where the wave function of the output optical state after the PINEM interaction with $\ket{\text{comb}(0)}$ ($\ket{\text{comb}(\pi)}$) is 
\begin{equation}
    \psi_{\pm}(x)=D(\pm g)\ket{\psi_{ph}}=\frac1{(\pi e^{-r})^{1/4}}\exp\left[-\frac{(x\mp g)^2}{2 e^{-r}}\right].
\end{equation}
Assume the initial electron state to be $\alpha \ket{\text{comb}(0)} +\beta\ket{\text{comb}(\pi)}$. If the amplitude-quadrature $X$ of the photon is measured to be $x$, the conditional state after the measurement is proportional to
\begin{equation}
    \alpha \psi_+(x)\ket{\text{comb}(0)}+\beta \psi_-(x)\ket{\text{comb}(\pi)}.
\end{equation}
In case of large $g e^{r/2}$, we have $\psi_+(x)\approx0$ for $x<0$ and we have $\psi_-(x)\approx0$ for $x>0$. For simplicity, we coarse-grain the measurement of X to a projective measurement onto the subspaces where $X > 0$ and $X < 0$. We consider the case for $X>0$ is measured, then the conditional electron-photon state is 
\begin{equation}
    \int_0^{+\infty}  dx \left(\alpha \psi_+(x)\ket{\text{comb}(0)}+\beta \psi_-(x)\ket{\text{comb}(\pi)}\right)\ket x.
\end{equation}
Thus, the fidelity for $\ket {\text{comb}(0) } $ is (condition on $X>0$)
\begin{equation}
    F(\ket{\text{comb}(0)}|X>0) =\frac{|\alpha|^2\int_0^\infty |\psi_+|^2 dx}{|\beta|^2\int_0^\infty |\psi_-|^2 dx+|\alpha|^2\int_0^\infty |\psi_+|^2 dx}.
\end{equation}
 The probability for $X>0$ is \begin{equation}
     P(X>0)= |\beta|^2\int_0^\infty |\psi_-|^2 dx+|\alpha|^2\int_0^\infty |\psi_+|^2 dx.
 \end{equation}
Similarly, we have 
\begin{align}
    F(\ket{\text{comb}(\pi)}|X<0) =\frac{|\beta|^2\int_{-\infty}^0 |\psi_-|^2 dx}{|\beta|^2\int_{-\infty}^0 |\psi_-|^2 dx+|\alpha|^2\int_{-\infty}^0 |\psi_+|^2 dx}.
    \\
     P(X<0)= |\beta|^2\int_{-\infty}^0 |\psi_-|^2 dx+|\alpha|^2\int_{-\infty}^0 |\psi_+|^2 dx.
\end{align}
Therefore, we can define the average fidelity for different measurement results
\begin{align}
     \bar F&=P(X>0) F(\ket{\text{comb}(0)}|X>0)+P(X<0) F(\ket{\text{comb}(\pi)}|X<0)
     \\
     &=|\alpha|^2\int_0^\infty |\psi_+|^2 dx+|\beta|^2\int_{-\infty}^0 |\psi_-|^2 dx
     \\
     &=\int_0^\infty |\psi_+|^2 dx
     \\
     &=\frac 12\text{erfc}( g e^{r/2}).
\end{align}
   For $ge^{r/2}>1.45$, $\bar F>98\%$.

\section{Quantum error correction after mitigation}
In our error mitigation scheme for the free electron qubit, a measurement outcome $\phi\ne 0$ projects the electron state to the subspace $\{\ket{\text{comb}(\phi)},\ket{\text{comb}(\phi+\pi)} \}$, thus deviating from the original code space. To correct such errors, we consider the work circuit of the free electron qubit, where the effect of the free electron can be expressed as a function of $g_wb$, e.g. $f(g_wb)$, with $g_w$ denoting the coupling strength for the free electron in the work circuit. In the original code space spanned by $\{\ket{\text{comb}(0)},\ket{\text{comb}(\pi)} \}$, the matrix elements of $f(g_wb)$ are
\begin{align}
    \bra{{\text{comb}(0)}} f(g_w b)\ket{{\text{comb}(0)}}= f(g_w)&,\bra{{\text{comb}(\pi)}} f(g_w b)\ket{{\text{comb}(\pi)}}= f(-g_w).
\end{align}
Whereas, in the projected subspace $\{\ket{\text{comb}(\phi)},\ket{\text{comb}(\phi+\pi)} \}$, the matrix elements become
\begin{align}
    \bra{{\text{comb}(\phi)}} f(g_w b)\ket{{\text{comb}(\phi)}}= f(g_w e^{i\phi})&,\bra{{\text{comb}(\phi+\pi)}} f(g_w b)\ket{{\text{comb}(\phi+\pi)}}= f(-g_w e^{i\phi}).
\end{align}
Therefore, the error can be corrected by updating $g_w$ to $g_w e^{-i\phi}$ according to the measured $\phi$, so that the matrix element of $f(g_wb)$ remains unchanged.

\section{EPR steering criteria}
Based on the conventional EELS and our proposed FECBPM, incompatible measurements can be performed on electron-photon states. Therefore, we have the ability to detect EPR steering between the electron and the optical mode according to Reid's criterion~\cite{reid1989demonstration}.

\subsection{EPR steering for the case where electron qubits interact with optical coherent states}
Here we will provide the analytical results of the first example, where an electron qubit interacts with an optical coherent state via the PINEM system. In this case, quadratures are measured for the optical mode, while Pauli operators are measured for the electron qubits.

Here, we will detect the EPR steering from electron to photons, that is, to evaluate the inferred variance of the optical mode with the measurement outcomes of the electron qubit. With direct calculation, the conditional states for $\sigma_x=\pm 1$ are obtained as $D(\pm g)\ket{i\alpha}$ with the probability being $P(\sigma_x=1)=P(\sigma_{x}=-1)=\frac 12$. Hence, the inferred variance of the amplitude-quadrature is $\Delta^2_{inf} X_A=\frac 12$, which is equal to that of coherent states. At the same time, the conditional states for $\sigma_z=\pm 1$ are $\frac 1{\sqrt N_{\pm}} (D(g)\ket {i\alpha}\pm D(-g)\ket{i\alpha})$ with the probability of $P(\sigma_z=\pm 1)=\frac 12(1\pm e^{-2|g|^2}\cos(4\alpha g))$. The expectation values of the phase-quadrature then becomes
\begin{equation}
    \langle P_A\rangle_{\pm}=\frac{\sqrt2 \alpha\pm\sqrt{2}e^{-2|g|^2}Re(e^{-4i\alpha g}(\alpha-ig))}{1\pm e^{-2|g|^2}\cos(4\alpha g)}.
\end{equation}
And the variances are
\begin{equation}
    \Delta^2P_{A\pm}=\frac{4\alpha^2+1\pm Re(e^{-2|g|^2 } e^{-4i\alpha g} (1-(g+i\alpha)^2))}{2(1\pm e^{-2|g|^2}\cos(4\alpha g))}- \langle P_A\rangle_{\pm}^2.
\end{equation}
Thus, the inferred variance can be calculated as
\begin{equation}
  \Delta^2_{inf}P_A= \frac 1{2-\frac{8g^2\sin^2(4\alpha g)}{-e^{4g^2}+\cos^2(4\alpha g)+4g^2\sin^2(4\alpha g)}}.
\end{equation}
It's directly checked that $e^{4g^2}>1+4g^2>\cos^2(4\alpha g)+4g^2\sin^2(4\alpha g)$, which indicates $\Delta^2_{inf} P<1/2$. Therefore, in the ideal case where the electron qubit interacts with optical coherent states with dissipation and thermal noise ignored, we find that the electron qubit can always steer the optical mode, $\Delta^2_{inf} P\Delta^2_{inf} X<1/4$, no matter what values of $\alpha$ and $g$ are chosen.

\subsection{EPR steering for where free electron interacts with optical thermal states}
In the second example where the free electron interacts with optical thermal states, we choose $X_A$ and $P_A$ as the optical number $N_{ph}$ and phase operators $\phi_{ph}$, respectively, which satisfy $[N_{ph},\phi_{ph}]=i$. Meanwhile, $X_B$ and $P_B$ are chosen as the electron energy $X_B=E_e$ and phase operators $P_B=\phi_e$, respectively. 

We first consider the special case where the initial state is $\ket{\psi_0}=\ket 0_{ph}\otimes\ket 0_e$, i.e. the initial optical state is prepared as a zero-temperature vacuum state, so that the state after a PINEM interaction becomes $\ket{\psi}=\sum c_k \ket k_{ph}\ket{-k} _e$. Thus, the conditional state after electron energy measurement is the optical Fock state with $\Delta^2N_{ph}=0$. And the conditional state after FECBPM is the optical coherent state, with $\Delta^2\phi_{ph}<+\infty$. Therefore, we have 
 \begin{equation}
    \Delta^2_{inf} N_{ph}\Delta^2_{inf} \phi_{ph}=0.
 \end{equation}
According to Reid's criterion, the EPR steering is detected if $\Delta^2_{inf}N\Delta^2_{inf}\phi<1/4$. This means that, in the ideal zero-temperature case, the electron can always steer the optical mode.

Then by involving practical conditions, the photon state is prepared as a thermal state $\rho_{ph}=\sum_{n=0}^{\infty} \frac{\bar{n}^{n}}{(\bar{n} + 1)^{n+1}} \, |n\rangle\langle n|$. Suppose that the electron state is prepared as the energy eigenstate $\rho_e=\ket E\bra E$, the final state after the PINEM interactions is expressed as $\rho=U(g)[\rho_e\otimes \rho_{ph}]U^\dagger(g)$. In order to evaluate the variance of phase $\phi$, we refer to the Susskind-Glogower operator $V=a_{ph}/\sqrt{a_{ph} a_{ph} ^\dagger }$~\cite{SG-operator}, so that $\Delta^2\phi$ is obtained as
\begin{equation}
    \Delta^2\phi=\frac{\Delta^2 C+\Delta^2S}{\langle C\rangle^2+\langle S\rangle^2},
\end{equation}
where $C=(V+V^\dagger)/2$ and $S=(V-V^\dagger)/(2i)$. These results have been used to produce Fig.~4(b) in the main text.


\bibliography{ref}